\documentclass[aps,prl,reprint,twocolumn,superscriptaddress,nofootinbib]{revtex4-1}

\usepackage{amssymb}
\usepackage{amsmath}
\usepackage{graphicx}
\usepackage{subfigure}
\usepackage{longtable}
\usepackage{color}
\usepackage{textcomp}
\usepackage{hyperref}

\begin{document}

\title{Correlated evolution of structure and mechanical loss of a sputtered silica film}

\date{\today}

\author{Massimo Granata}
	\email[]{m.granata@lma.in2p3.fr}
	\affiliation{Laboratoire des Mat\'eriaux Avanc\'es, CNRS/IN2P3, F-69622 Villeurbanne, France}	
\author{Elodie Coillet}
	\affiliation{Institut Lumi\`ere Mati\`ere, CNRS (UMR 5306), Universit\'e de Lyon, F-69622 Villeurbanne, France}
\author{Val\'erie Martinez}
	\email[]{valerie.martinez@univ-lyon1.fr}
	\affiliation{Institut Lumi\`ere Mati\`ere, CNRS (UMR 5306), Universit\'e de Lyon, F-69622 Villeurbanne, France}
\author{Vincent Dolique}
	\affiliation{Laboratoire des Mat\'eriaux Avanc\'es, CNRS/IN2P3, F-69622 Villeurbanne, France}
\author{Alex Amato}
	\affiliation{Laboratoire des Mat\'eriaux Avanc\'es, CNRS/IN2P3, F-69622 Villeurbanne, France}
\author{Maurizio Canepa}
	\affiliation{OPTMATLAB, Dipartimento di Fisica, Universit\`a di Genova, Via Dodecaneso 33, 16146 Genova, Italy}
	\affiliation{INFN, Sezione di Genova, Via Dodecaneso 33, 16146 Genova, Italy}
\author{J\'er\'emie Margueritat}
	\affiliation{Institut Lumi\`ere Mati\`ere, CNRS (UMR 5306), Universit\'e de Lyon, F-69622 Villeurbanne, France}
\author{Christine Martinet}
	\affiliation{Institut Lumi\`ere Mati\`ere, CNRS (UMR 5306), Universit\'e de Lyon, F-69622 Villeurbanne, France}
\author{Alain Mermet}
	\affiliation{Institut Lumi\`ere Mati\`ere, CNRS (UMR 5306), Universit\'e de Lyon, F-69622 Villeurbanne, France}
\author{Christophe Michel}
	\affiliation{Laboratoire des Mat\'eriaux Avanc\'es, CNRS/IN2P3, F-69622 Villeurbanne, France}
\author{Laurent Pinard}
	\affiliation{Laboratoire des Mat\'eriaux Avanc\'es, CNRS/IN2P3, F-69622 Villeurbanne, France}
\author{Beno\^it Sassolas}
	\affiliation{Laboratoire des Mat\'eriaux Avanc\'es, CNRS/IN2P3, F-69622 Villeurbanne, France}
\author{Gianpietro Cagnoli}
	\affiliation{Laboratoire des Mat\'eriaux Avanc\'es, CNRS/IN2P3, F-69622 Villeurbanne, France}
	\affiliation{Institut Lumi\`ere Mati\`ere, CNRS (UMR 5306), Universit\'e de Lyon, F-69622 Villeurbanne, France}

\begin{abstract}
Energy dissipation in amorphous coatings severely affects high-precision optical and quantum transducers. In order to isolate the source of coating loss, we performed an extensive study of Raman scattering and mechanical loss of a thermally-treated sputtered silica coating. Our results show that loss is correlated with the population of three-membered rings of Si-O$_4$ tetrahedral units, and support the evidence that thermal treatment reduces the density of metastable states separated by a characteristic energy of about 0.5 eV, in favour of an increase of the states separated by smaller activation energies. Finally, we conclude that three-fold rings are involved in the relaxation mechanisms only if they belong to more complex chain-like structures of 10 to 100 tetrahedra.
\end{abstract}

\pacs{xx.yy.zz, ii.jj.ll}


\maketitle

Thermal noise in amorphous dielectric materials is a fundamental limitation for a large number of precision experiments based on optical and quantum transducers, such as gravitational-wave detectors \cite{Adhikari14}, optomechanical resonators \cite{Aspelmeyer14}, frequency standards \cite{Kessler12}, quantum computers \cite{Martinis05} and atomic clocks \cite{Ludlow2015}. In these devices, thermally-driven random structural relaxations introduce decoherence in vibrational or electronic states. Any observable that is related to these states experiences fluctuations, i.e. noise. In mechanical experiments, this decoherence distributes the thermal energy of vibrations, stored in the normal modes, all over the frequency spectrum.

As stated by the fluctuation-dissipation theorem (FDT) \cite{Callen52}, the same structural relaxations give origin to energy dissipation: in linear systems at the thermodynamic equilibrium, the power spectral density of the fluctuations associated to any observable is proportional to the dissipative part of their dynamics. This fact is of the outmost importance, since in many cases the fluctuations are difficult to measure whereas the energy dissipation is in general more accessible. Dissipation of mechanical energy in materials is quantified by the so-called loss angle $\phi=\text{tan}^{-1}(\Im{(E)}/\Re{(E)})$, where $\Im{(E)}$ and $\Re{(E)}$ are the imaginary and real part of the Young’s modulus $E$, respectively. Although in principle there should be a loss angle associated to each elastic constant, we assume here that $\phi$ is the same for all of them; our measurements confirm this hypothesis. In the harmonic analysis, the imaginary part of the elastic constant is related to the retarded response of the material (anelasticity) \cite{Nowick72}.

In amorphous materials, the loss comes from unknown relaxation processes whose features are fairly well explained by the asymmetric double-well potential model \cite{Gilroy81}: the structure changes locally between two metastable states separated by a barrier of height $V$, and has a typical relaxation time $\tau \propto e^{V/k_B T}$. As a consequence, loss is a function of temperature, $\phi \equiv \phi(T)$. Remarkably, the temperature dependence is similar in many different bulk amorphous materials \cite{Topp96}, suggesting that this property might be determined by a set of universal laws. In amorphous coatings, on the contrary, the structure as well as $\phi(T)$ vary with the synthesis process.

Within this landscape, silica (SiO$_2$) shows an unusual and extremely interesting behavior, which makes it a suitable candidate to study the relaxation mechanisms of structural loss. In its bulk form, i.e. fused silica, the internal friction $\phi \sim 5 \cdot 10^{-9}$ is about 4 orders of magnitude lower than any other metal oxides at $T_0 \sim 300$ K \cite{Topp96, Ageev04}; yet, if deposited as a ion-beam sputtered (IBS) film, it shows a much higher damping -- $\phi \sim 5 \cdot 10^{-5}$ at $T_0$ \cite{Principe15, Granata16}. A change in the energy distribution of relaxation mechanisms is likely responsible of this behaviour, but, despite the fact that silica is one of the most studied glass in material science and that its $\phi(T)$ is nowadays very well characterized \cite{Penn06, Travasso07, Martin14}, these mechanisms are still unknown. In order to isolate such mechanisms, we performed an extensive study of Raman scattering and loss of a thermally-treated IBS silica coating, and we addressed the possible correlations between damping and structure. The structure of silica is a 3-dimensional network composed of SiO$_4$ tetrahedral units arranged in rings of assorted sizes, from 3 to 10 tetrahedra \cite{Jin94}. Raman spectroscopy is sensitive to the vibration of local structures in vitreous systems at short- to middle-range scale (few to 10 {\AA}), and particularly to the distribution of angles between tetrahedra and to the population of three-fold and four-fold rings.

We deposited the silica coating on a fused silica disk-shaped mechanical resonator ($\varnothing\,$ 3", 1 mm thick) for the characterization of loss, and on a fused silica tablet ($\varnothing\,$ 1", 5 mm thick) for the Raman study. Prior to deposition, the disk was annealed at 900$^{\circ}$ C for 10 hours and the tablet has been coated with 100 nm of tantalum (which has no active Raman vibrational modes) in order to acquire the signal of the silica coating only.
\begin{figure}
	\includegraphics[width=0.40\textwidth]{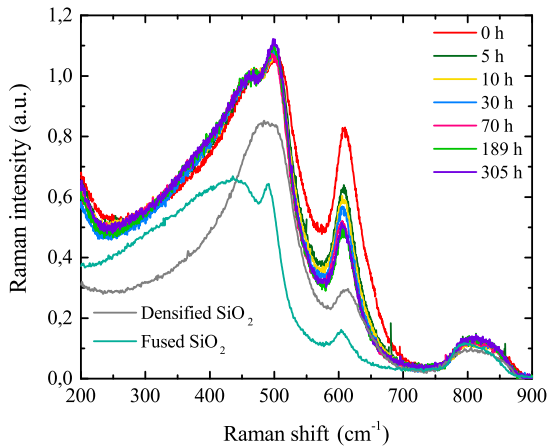}
	\caption{\label{FIGcompSp} (Color online) Raman spectra of SiO$_2$: comparison of the as-deposited coating ($t_a=0$ h) to fused silica, densified silica and thermally-treated coating with different cumulative annealing time $t_a$ (the spectra have been arbitrarily shifted along the ordinate axis for clarity).}
\end{figure}
3 $\mu$m of amorphous SiO$_2$ coating have been deposited on both sides of the disk and on the metal-coated surface of the tablet, in a Veeco SPECTOR dual-IBS chamber. Argon and oxygen were fed into both sources, with flow rates of 18 sccm and 15 sccm in the sputtering source and 3 sccm and 12 sccm in the assistance source, respectively. During the coating process, the energy of the coating particles impinging on the substrate was of the order of 10 eV, and the temperature of the substrates was approximately $80^{\circ}$ C. After coating, both samples have been repeatedly annealed together at 500$^{\circ}$ C, a temperature much lower than the silica glass-transition temperature $T_g\sim 1200^{\circ}$ C, for progressively increasing time. 500$^{\circ}$ C is a typical annealing temperature for optical coatings \cite{Anghinolfi13, Granata16}. The Raman spectrum and the coating loss have been measured at room temperature after each annealing step, to follow at the same time the evolution of structure and damping.

Raman spectra were recorded with a LabRAM HR Evolution micro-Raman spectrometer, equipped with three super-notch filters to attenuate the Rayleigh line and with a Peltier-cooled charge-coupled device. The incident light was emitted by a YAG:Nd$^{3+}$ laser at 532 nm. Spectra were recorded from 200 to 900 cm$^{-1}$ with a x100 objective delivering 6 mW on the sample, avoiding heating or sample damage. To measure the coating loss, we applied the resonance method \cite{Nowick72} to the disk, measuring the ring-down time $\tau$ of 9 vibrational modes from 1 to 17 kHz. For each mode of frequency $f_n$, the loss $\phi_n=\pi f_n\tau_n$ is the linear combination of the losses  of substrate and coating, where the coefficients are the fraction of elastic energy stored in each part. The double measurement of loss before and after the coating deposition allows the estimation of the coating loss only \cite{Granata16}. A clamping-free system named Gentle Nodal Suspension (GeNS) \cite{Cesarini09} has been used to suspend the disk, as it highly suppresses the systematic damping due to suspension and allows a high reproducibility of the results \cite{Granata15}. More details about our experimental setup can be found elsewhere \cite{Granata16}.

\begin{figure}
 \subfigure[]{
	\includegraphics[width=0.40\textwidth]{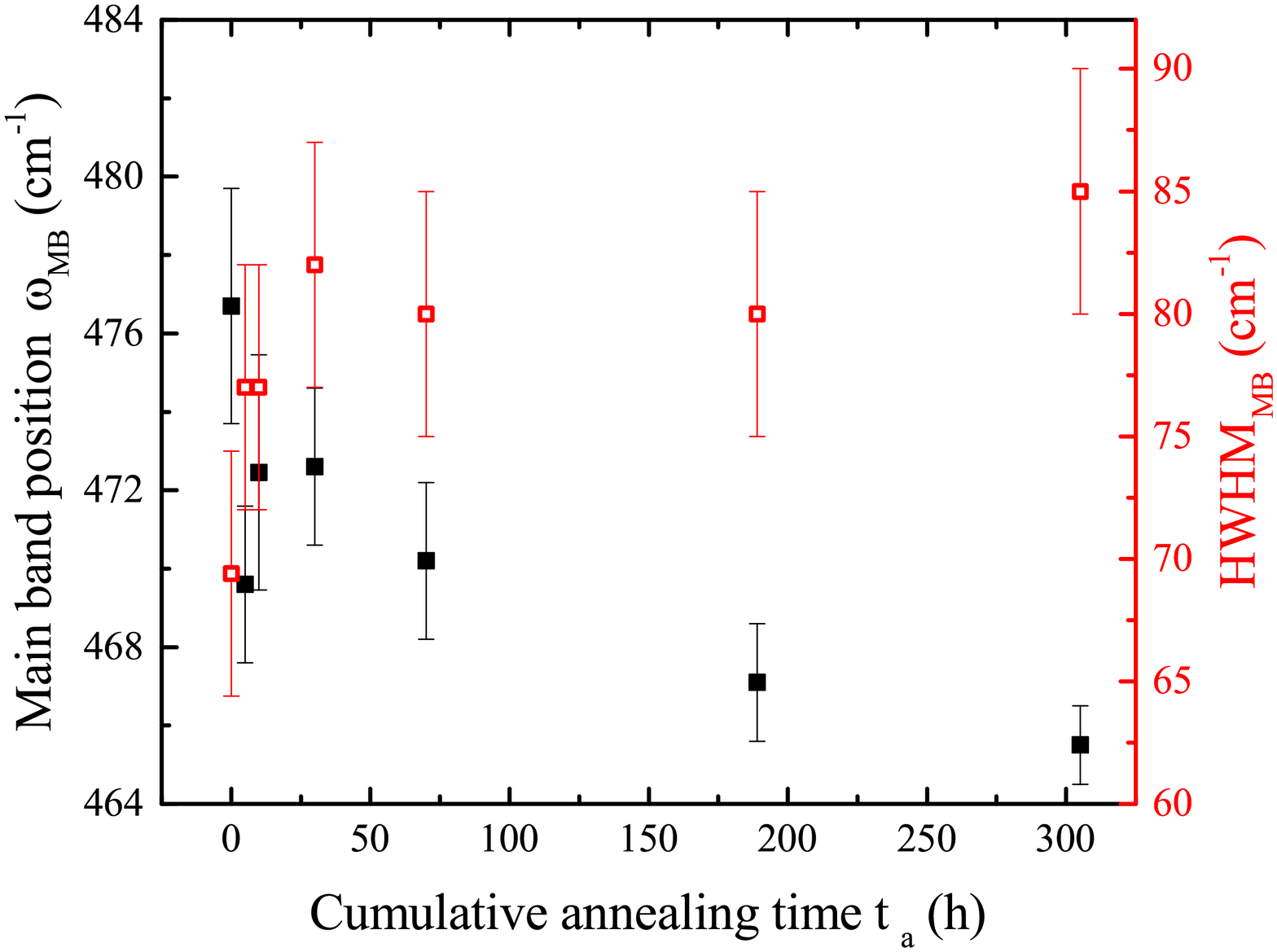}
	\label{FIGmainBand}
 }
 \subfigure[]{
   \includegraphics[width=0.40\textwidth]{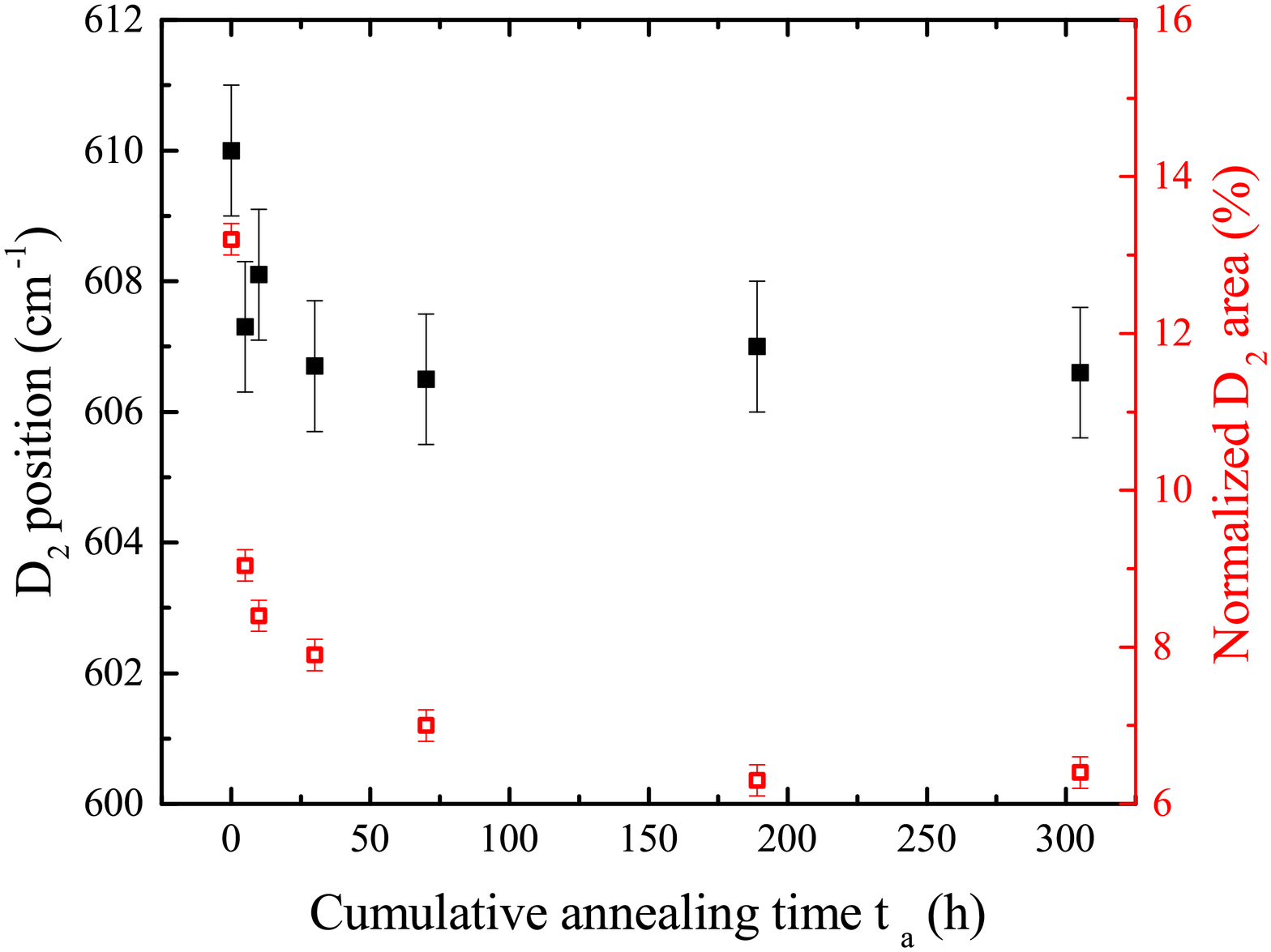}
   \label{FIGd2}
 }
	\caption{(Color online) Evolution of the film spectral features as a function of the cumulative annealing time $t_a$: a) $\omega_{\text{mb}}$ and HWHM$_{\text{mb}}$; b) D$_2$ normalized area and frequency.}
\end{figure}

Fig.\ref{FIGcompSp} shows that the Raman spectra of our film is more similar to that of densified silica than to that of fused silica. The density $\rho$ of our film can be inferred from its correlation with the half width at half maximum of the main band (HWHM$_{\text{MB}}$), established for fused silica samples that underwent a cycle of densifications obtained after a hydrostatic high-pressure cycle (up to 26 GPa and 2.66 g/cm$^3$) at room temperature \cite{Martinet15}. Using this curve and measuring HWHM$_{\text{MB}}$ for our sputtered silica (Fig.\ref{FIGmainBand}), we obtain $\rho = 2.37 \pm 0.02$ g/cm$^3$. This is in agreement with the independent estimate $\rho= 2.33 \pm 0.04$ g/cm$^3$, obtained through the direct measurement of coating mass (with a balance) and through the analysis of spectroscopic-ellipsometry measurements (with a J. A. Woollam VASE instrument) yielding coating thickness and refraction index. The densification can be inferred also from the frequency $\omega_{\text{MB}}$ of the main band maximum, which corresponds to the bending mode of bridging oxygen (Si-O-Si). $\omega_{\text{MB}}$ is directly linked to the mean inter-tetrahedral angle $\theta$ via the Sen and Thorpe central-force model \cite{Sen77},
\begin{equation}
\label{EQsenThorpe}
\omega_{\text{mb}} = (2 \frac{\alpha}{m_{\text{O}}})^{\frac{1}{2}} cos(\frac{\theta}{2}) \ ,
\end{equation}
where $\alpha = 1.622 \cdot 10^{-7}$ g/mol/cm$^2$ is the restoring constant central force between Si and O atoms, $m_{\text{O}} = 16$ g/mol is the oxygen mass. For fused silica $\theta \sim 144^{\circ}$, corresponding to $\omega_{\text{MB}} = 434.7 \pm 0.5$ cm$^{-1}$, and this value decreases with the densification ratio $\Delta\rho/\rho$ \cite{Martinet15}. For our coating $\omega_{\text{MB}} = 476.7 \pm 0.5$ cm$^{-1}$, thus $\theta \sim 140.9^{\circ}$.

In fused silica, the ring statistic –- characteristic of medium range order of silicate glasses -– is peaked on six-fold rings \cite{Jin94, Huang04}, and is shifted towards smaller rings when density increases via cold compression \cite{Jin94} beyond the elastic limit. The two sharp bands at 490 cm$^{-1}$ (hereafter {\it $D_1$}) and at 605 cm$^{-1}$ ($D_2$) are assigned to four-fold and three-fold ring breathing modes, respectively, and it has been shown \cite{Pasquarello98, Burgin08} that the normalized $D_2$ area is related to the three-fold ring population. From Fig.\ref{FIGcompSp}, we observe that in the coating spectrum both $D_1$ and $D_2$ are more intense than in the spectra of both fused silica and densified silica. Its $D_2$ area (normalized to the area from 230 to 700 cm$^{-1}$) is higher than that of both densified and fused silica (of about 6 times compared to the latter). Finally, the $D_2$ position in our film is shifted towards higher frequencies with respect to that of fused silica (Fig.\ref{FIGd2}): this finding indicates the presence of an internal stress. Following Ref.\cite{Hehlen09}, the $D_2$ shift of 5 cm$^{-1}$ corresponds to a 0.4$^{\circ}$ decrease of the Si-O-Si angles due to stress-induced ring puckering \cite{Brinker88, Barrio93}.

Fig.\ref{FIGcompSp} also shows the Raman spectra of the annealed coating for different cumulative annealing time $t_a$, increasing from 5 to 300 hours, compared to the as-deposited ($t_a = 0$ h) coating spectrum. The evolution of the coating spectral features as a function of $t_a$ is shown in Figg.\ref{FIGmainBand} and \ref{FIGd2}.
$\omega_{\text{MB}}$ shifts from $476.7 \pm 0.5$ cm$^{-1}$ at $t_a = 0$ h to $465.5 \pm 0.5$ cm$^{-1}$ at $t_a = 300$ h, without reaching a plateau. This corresponds to an increase $\Delta\theta = 0.9^{\circ}$, which is coherent with FTIR measurements of sputtered silica coatings showing that $\theta$ increases with annealing \cite{Hirose06}, and can be related to a less dense structure. HWHM$_{\text{MB}}$ increases along with $t_a$, denoting a widening of the $\theta$ distribution: HWHM$_{\text{MB}}$ is $69 \pm 5$ cm$^{-1}$ at $t_a=0$ and $85 \pm 5$ cm$^{-1}$ at $t_a = 300$ h. For comparison, HWHM$_{\text{mb}} \sim 120 \pm 10$ cm$^{-1}$ for fused silica. Thus the annealed film is less homogeneous in terms of $\theta$ values than the as-deposited film.

The normalized $D_2$ area decreases monotonically, following a stretched exponential law \cite{Hirose06} with $\tau = 6.3 \pm 1.1$ h and $\beta = 0.36 \pm 0.05$. This relaxation time is relatively short: magnetron-sputtered silica deposited at 520$^{\circ}$ C has the same relaxation time at about 800$^{\circ}$ C, and follows the Arrhenius law with activation energy of $5.4 \pm 0.2$ eV \cite{Hirose06}. Considering also that relaxation times of fused silica (with fictive temperature $T_f = 1100^{\circ}$ C) are $6\cdot10^4$ times longer \cite{Hirose06}, we might say that the deposition temperature is more effective than the annealing temperature in stabilizing sputtered silica. Unfortunately, those data are not conclusive since relaxations in \cite{Hirose06} were observed in the infra-red reflectivity (stretching mode of Si-O-Si) whereas our data concern the $D_2$ evolution, and the fact that in \cite{Hirose06} $\beta < 0.3$ indicates the presence of a broad spectrum of relaxation mechanisms.

It is more difficult to quantify the evolution of the four-fold ring population, since the $D_1$ band is partly overlapping with the main band and their deconvolution is not trivial. Qualitatively, $D_1$ population increases with $t_a$.
 
All the above measurements lead to explain the structural evolution of our silica coating by stress relaxation. It is worth clarifying that such stress is internal and cannot be removed if the film is detached from the substrate. We propose here that the internal stress accelerates the relaxations. This can justify why our low-temperature-deposited silica relaxes faster than a high-temperature-deposited one, and still faster than fused silica \cite{Hirose06} (we can assume that this latter is free of stress). Our IBS silica reached a plateau, but this does not correspond to a stress-free structure.

\begin{figure}
 \subfigure[]{
	\includegraphics[width=0.40\textwidth]{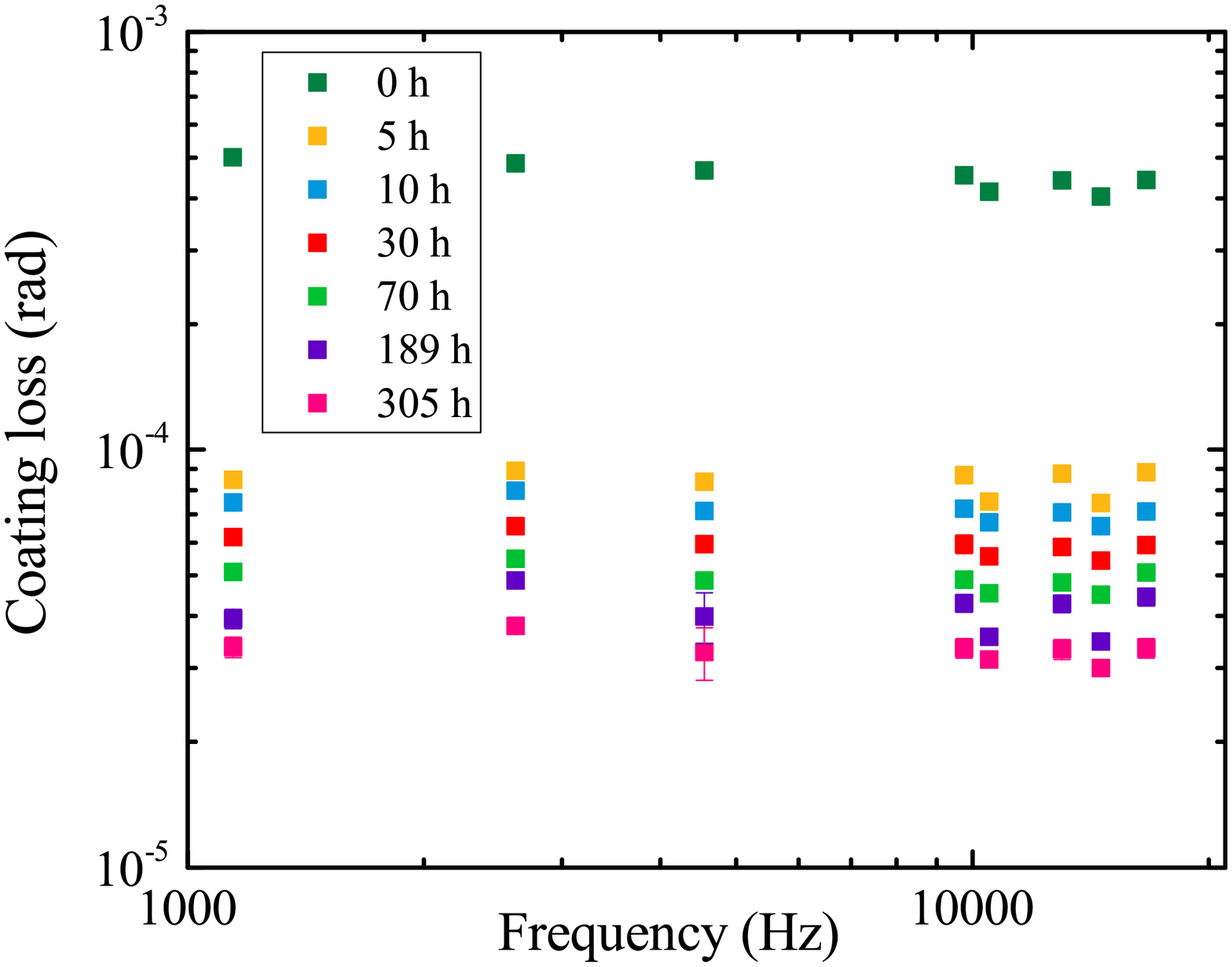}
	\label{FIGlossVSf}
 }
 \subfigure[]{
   \includegraphics[width=0.40\textwidth]{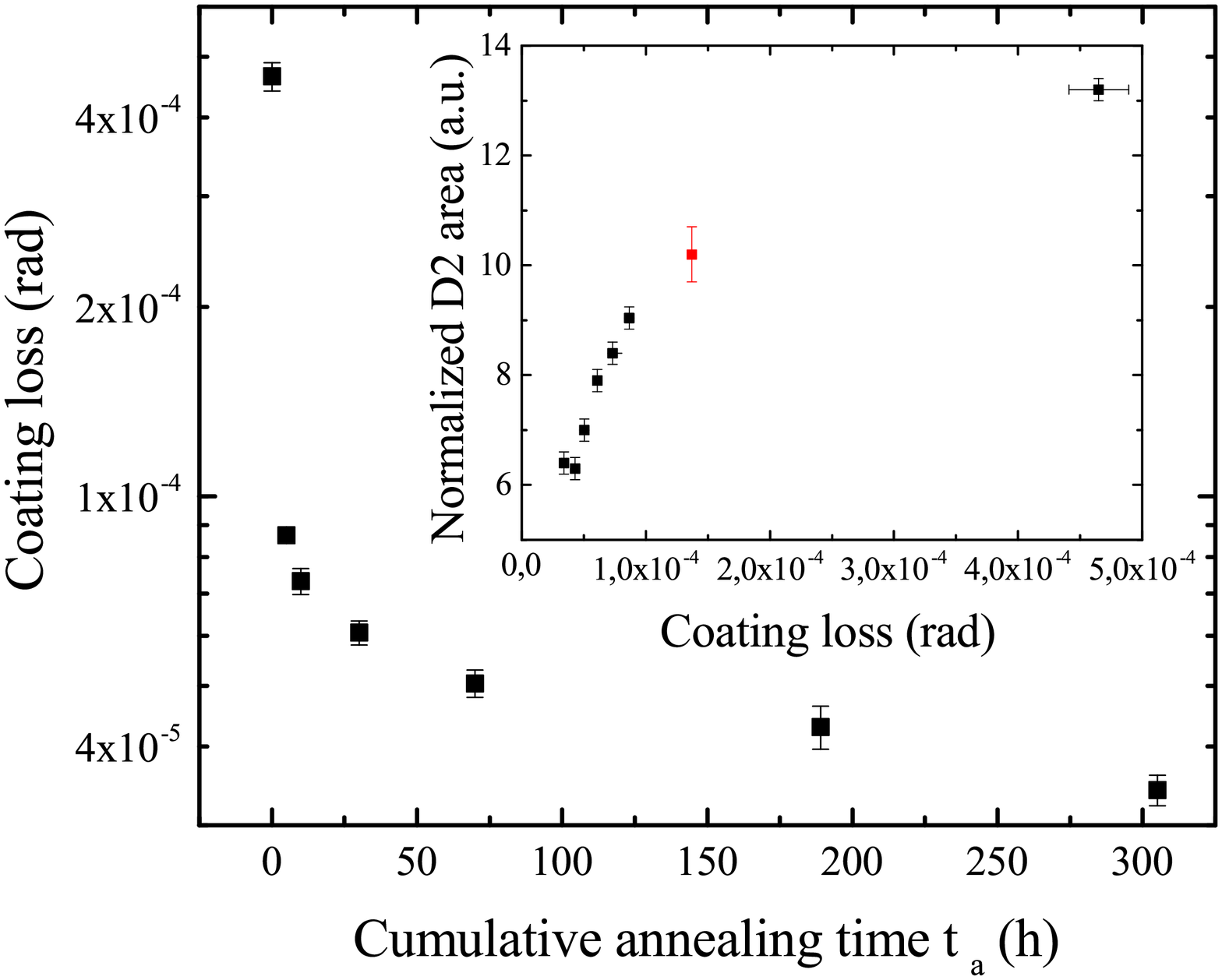}
   \label{FIGavLossVSannTime}
 }
	\caption{(Color online) Results of the mechanical characterization: a) coating loss as a function of frequency, for different values of the cumulative annealing time $t_a$; b) frequency-averaged coating loss $\langle \phi \rangle_{\omega}$ as a function of $t_a$ and correlation between normalized $D_2$ area and $\langle \phi \rangle_{\omega}$ (the red point is from another IBS silica film with different deposition parameters).}
\end{figure}

Our loss measurements are shown on Fig.\ref{FIGlossVSf}, where each data series corresponds to a given value of $t_a$. The heat treatment progressively reduces the loss, the most dramatic decrease happening for $t_a = 5$ h. For the outliers at 10.5 and 14.6 kHz, the estimated coating energy fraction is different from the computed one; as this is not explained so far, we excluded them from the computation of the frequency-averaged coating loss $\langle \phi \rangle_{\omega}$ presented in Fig.\ref{FIGavLossVSannTime} (as they would only induce an offset which would not change our analysis).
The observed loss evolution is due to the coating only: the substrate being annealed at 900$^{\circ}$ C before deposition, no variation of loss has been measured after the post-deposition annealing at 500$^{\circ}$. The correlation between the normalized $D_2$ area and $\langle \phi \rangle_{\omega}$ is shown in the inset of Fig.\ref{FIGavLossVSannTime}. Clearly, $\langle \phi \rangle_{\omega}$ increases monotonically with the area underlying the $D_2$ line, and thus with the three-fold ring population. Fig.\ref{FIGavLossVSannTime} also shows a single data point relative to another IBS silica film (not annealed) deposited with a different set of parameters, and the correlation still holds. Recent molecular-dynamics simulations \cite{Hamdan14} established that structural relaxation comes from the twisting of chains of few tens of SiO4 tetrahedra, via rotation and stretching of Si-O bonds; between 10 and $10^2$ atoms are involved in this reorganization, but an investigation on the ring population inside these relaxing chains has not yet been carried out at present.

Relaxation mechanisms active at room temperature and at acoustic frequencies should have barrier heights of about 0.5 eV \cite{Gilroy81}. The activation energy of 3-membered rings has been measured to be about 0.43 eV \cite{Barrio93, Shimodaira06}. The similarity between these energies could explain why loss and $D_2$ ring population are correlated: the annealing reduces the density of metastable states separated by about 0.5 eV, in favour of an increase of the states separated by smaller activation energies. This increase is suggested by our measurements, as shown by the increase of the $D_1$ ring population, and by low-temperature ($T<100$ K) loss measurements \cite{Weiss96}. These latter measurements show that densified silica has less internal friction than fused silica at 12 kHz. Fused silica shows a loss peak at 30 K compatible with a distribution of barrier heights centered around 50 meV.

It is not clear whether the three-fold rings are one of the metastable states involved in the relaxations that cause the loss in IBS silica coatings. The relaxation time of our IBS silica, as measured from the loss data, is $5 \pm 2$ minutes ($\beta = 0.18 \pm 0.02$); loss in fused silica is several orders of magnitude lower than in our coating, despite the amplitude of its $D_2$ line (Fig.\ref{FIGcompSp}). These facts suggest that metastable structures relevant for loss are different from three-fold rings.

In summary, IBS deposition leads to a dense and internally-stressed silica film. This outcome is consistent with previous studies showing that IBS and ion-assisted deposition yield highly-densified silica coatings \cite{Hirose06, Weiss96, Martin83, Sainty84, Bhumbra95}. Thanks to internal stress, IBS silica relaxes towards a less dense structure even if $T \ll T_g$. The loss of the silica coating is correlated with the population of three-fold rings, and this correlation holds even when the coating structure is determined by different deposition parameters. three-fold rings are involved in the relaxation mechanisms only if they belong to more complex structures.

The authors gratefully acknowledge support from the LabEx Lyon Institute of Origins (LIO, grant ANR-10-LABX-0066) within the program `Investissements d'Avenir' (grant ANR-11-IDEX-0007) of the French National Research Agency (ANR), and would like to thank the CECOMO Platform of vibrational spectroscopy.

\end{document}